\documentstyle[eqsecnum,aps,psfig,twocolumn]{revtex}

\begin{document}

\title{\large \bf EXTERNAL FLUCTUATIONS IN A PATTERN-FORMING
INSTABILITY
}
\author{\large
J. Garc\'{\i}a-Ojalvo $^*$}
\address{School of Physics,
Georgia Institute of Technology, Atlanta GA 30332-0435}
\author{
and \\ \large J.M. Sancho}
\address{
Departament d'Estructura i Constituents de la
Mat\`{e}ria, Facultat de F\'{\i}sica
\\ Universitat de Barcelona,
Diagonal 647, E-08028 Barcelona, Spain}
\maketitle
\begin{abstract}
The effect of external fluctuations on the formation of spatial
patterns is analysed by means of a stochastic Swift-Hohenberg
model
with multiplicative space-correlated noise. Numerical simulations
in two dimensions show a shift of the bifurcation point
controlled by the intensity of the multiplicative noise. This
shift takes place in the ordering direction (i.e. produces
patterns), but its magnitude decreases with that of the noise
correlation length. Analytical arguments are presented to explain
these facts.
\end{abstract}

\vskip5mm

\narrowtext

\section{Introduction}
\label{sec:introduction}
A large amount of experimental spatially-extended systems
exhibiting non-equilibrium transitions controlled by the
environment exist (for a review, see \cite{cross93}). These
transitions usually occur in such a way that the system
departs from an initial homogeneous state when an external
control parameter surpasses a certain threshold value.
Frequently, this instability leads to the appearance of
non-equilibrium spatiotemporal dissipative structures, which
subsist as long as the external driving stress which has produced
the transition persists \cite{manneville90}.
This is the case, for instance, in the hydrodynamic convective
structures controlled by the temperature difference between the
plates of a Rayleigh-B\'enard cell \cite{chandras61,meyer87}, or
in the appearance of transverse structures in a laser beam
controlled by the pumping rate \cite{brambilla94,lega94a}.
Usually, most of the features present in these pattern-forming
processes are satisfactorily captured by a model equation which was
introduced by Swift and Hohenberg
\cite{swift77} with the initial objective of describing the
effect of fluctuations at the onset of convection in the
Rayleigh-B\'enard cell mentioned above. Since then, the
Swift-Hohenberg (SH) equation (and proper modifications of it
\cite{xi94,hilali95}) has proved its great usefulness in this
field, not only in a hydrodynamical context
\cite{mandel93,lega94b}. The equation reads:
\begin{equation}
\label{eq:ash}
\frac{\partial u\left(\vec{x},t\right)}{\partial t}=r\,u
-\left[\nabla^2+k_0^2\right]^2 u-u^3+\eta\left(\vec{x},t\right)
\end{equation}
Here $r$ is the bifurcation parameter (the temperature gradient in
the Rayleigh-B\'enard case, or the pumping rate in the laser
case), which controls the appearance of a pattern
of characteristic length of order $k_0^{-1}$.
The field $\eta$ is a random function of space and time,
representing internal noise of the system. Usually this term is
statistically described by a gaussian probability distribution
with zero mean and correlation
\begin{equation}
\label{eq:int-noise}
\left< \eta\left(\vec{x},t\right)\;\eta\left(\vec{x}\,',t'\right)
\right>\;=
\;2\,\varepsilon\,\delta\left(\vec{x}-\vec{x}\,'\right)\,\delta\left(t-t'
\right)
\end{equation}
This white noise accounts for hydrodynamic thermal fluctuations
in the convective instability, or for spontaneous
emission in the optical case. Its (dimensionless)
strength $\varepsilon$ is very small, so it has no qualitative effect on
the pitchfork bifurcation exhibited by this model.

External fluctuations, on the other hand, are known to have
non-trivial qualitative effects on the behaviour of non-linear
dynamical systems. In particular, zero-dimensional systems (i.e.
systems with no spatial dependence) have been known for more than
a decade to exhibit transitions {\em induced} by external noise
(see \cite{horsthemke84} for an extensive
review). In the last few years, interest on studying the
influence of external fluctuations on spatially-extended systems
has grown \cite{doering89,elder92,ojalvo93,becker94,broeck94}.
This influence is likely to be more relevant in these systems
than in the previous homogeneous ones, due to the existence of
symmetry-breaking effects in the transitions occurring in them.

In the particular case of the SH model, Elder
{\em et al.} \cite{elder92} observed a disordering effect of the
additive noise present in Eq. (\ref{eq:ash}), when it is no
longer considered to be internal, so that its intensity $\varepsilon$ can
be arbitrarily large. Hence, the stripe structure (roll structure,
in the hydrodynamical terminology) suffered a transition from a
smectic (large regions of parallel stripes) to an isotropic
(disordered system with short-range order) regime as the
intensity of the additive external noise increased.

Yet another possible (even more reasonable) source of external noise
exists, namely the control parameter of the bifurcation. The fact
that this parameter is directly related to an external constraint
imposed by the observer makes it likely to be affected by fluctuations.
Let us now denote by $r$ the mean value of this fluctuating
control parameter. Now the SH equation reads:
\begin{equation}
\label{eq:msh}
\frac{\partial u\left(\vec{x},t\right)}{\partial t}=
\left[r+\xi\left(\vec{x},t\right)\right]\,u
-\left[\nabla^2+k_0^2\right]^2 u-u^3+\eta\left(\vec{x},t\right)
\end{equation}

Hence the random field $\xi$ is a zero-mean external multiplicative
noise (chosen gaussian) whose correlation should not in principle be
assumed to be white. We shall choose here a noise colored in
space and white in time:
\begin{equation}
\label{eq:ext-noise}
\left< \xi\left(\vec{x},t\right)\;\xi\left(\vec{x}\,',t'\right)
\right>\;= \;2\,D\left(\frac{\vec{x}-
\vec{x}\,'}{\lambda}\right)\,\delta\left(t-t' \right)
\end{equation}
$\lambda$ is the correlation length of the noise. This model was
studied in Ref. \cite{ojalvo93} in the limit case
$\lambda \longrightarrow 0$ (noise white also in space) and
nontrivial effects were indeed found on the original bifurcation,
which was shifted into the pattern region as the intensity of the
external noise increased. Hence the role of this external noise
is, at least for some values of its intensity,
opposite to that of the additive case. A linear stability
analysis of the first and second moments of the relevant variable
of the system leads to analytical results which are in agreement
with simulations \cite{ojalvo93,becker94}. Now we are
interested in the influence
of the correlation length of the noise (which in some other
problems is known to reduce the effective value of the noise
intensity \cite{ojalvo92a,ojalvo94a}). The mathematical treatment of the
Fokker-Planck equation to study spatially-extended systems with multiplicative
colored noise is also presented.

This paper is organised as follows. Next Section presents the
results obtained by numerical simulation of the model. A linear
stability analysis of the structure function of the system is
made in Section \ref{sec:sth}, where a comparison with the
previous numerical results shows good agreement. Section
\ref{sec:th} contains a generalisation of the linear stability
analysis treatment to higher-order statistical moments.
Finally some conclusions are stated. An Appendix contains some
details on the Fokker-Planck approach to the problem.

\section{Numerical analysis of the full nonlinear model}
\label{sec:sim}

The non-equilibrium, nonlinear stochastic problem presented in the
previous Section does not admit an exact analytical study. Hence
our first approach to the problem is a numerical one, and
constitutes the best way of obtaining a first insight into the
effects of external fluctuations in the pattern-forming
instability developed by the SH model.

The behaviour of the model in the absence of multiplicative noise
is well known. When $\varepsilon$ is small,
the homogeneous state $u=0$ is stable for a
negative value of the external control parameter $r$. For $r\simeq 0$,
a bifurcation takes place from this homogeneous situation to an
inhomogeneous state composed of stripes. The influence of an
uncorrelated fluctuating control parameter
in this non-equilibrium spatially-extended transition
was preliminarly analysed in Ref. \cite{ojalvo93}, revealing a
nontrivial ordering effect. Here we extend the numerical analysis
to a colored case.

\subsection{Algorithm}

We begin by discretizing space in a regular two-dimensional
square lattice with $L\times L$ cells of size $\Delta x$. Now the
SH model can be written in the following general form:
\begin{equation}
\label{eq:dgshe}
\frac{\partial u_i}{\partial t} =
f_i\left(u\right)+g_{ij}\left(u\right)\xi_j(t)+
\eta_i(t)
\end{equation}
with
\begin{equation}
\label{eq:dnc-int}
<\eta_i(t)\,\eta_j(t')>\;=\;2\,\frac{\varepsilon}{\Delta x^2}\;
\delta_{ij} \,\delta(t-t')
\end{equation}
\begin{equation}
\label{eq:dnc-ext}
<\xi_i(t)\,\xi_j(t')>\;=\;2\; D_{i-j}\;
\delta(t-t')\;.
\end{equation}
Cells are named with one index and repeated indexes are summed up.
The deterministic force $f$ and the coupling function $g$
are in our particular case:
\begin{equation}
\label{eq:df-f}
f_i\left(u\right)\;=\;r\;u_i-\left(\nabla^4 u
\right)_i-2\left(\nabla^2u\right)_i-u_i-u_i^3
\end{equation}
\begin{equation}
\label{eq:cf-gh}
g_{ij}\left(u\right)\;=\;u_i\;\delta_{ij}
\end{equation}
The discrete Laplacian operator is defined as
\begin{equation}
\left(\nabla^2 u\right)_i= \frac{1}{(\Delta x)^2}
\sum_n \left(u_n-u_i\right)
\label{eq:laplaced}
\end{equation}
where the sum extends over the set of nearest neighbours of site i.
Now dynamical evolution is discretized in time (let $\Delta t$
be the time integration step) and an algorithm can be developed
from standard techniques \cite{sancho82a}. Up to ${\cal O}
\left(\Delta t^{3/2}\right)$, it reads
\begin{eqnarray}
u_i(t+\Delta t)=u_i(t)+f_i\left(u(t)\right)\Delta t+
g_{ij}\left(u(t)\right)X_j
\nonumber \\
+Y_i+\frac{1}{2}\;\frac{\partial g_{ij}}{\partial u_k}\;g_{kl}\;X_j\;X_l
+\frac{1}{2}\;\frac{\partial g_{ij}}{\partial u_k}\;X_j\;Y_k
\label{eq:dgshe-alg}
\end{eqnarray}
where $Y_j$ is a gaussianly distributed random number with zero mean and
variance $2\,\Delta t \,\varepsilon/\Delta x^d$ to be placed at site j.
$X_j=\sqrt{2\,\Delta t}\;Z_j$, where $Z_j$ is a random field
correlated in space, whose generation procedure will be
described later. According to (\ref{eq:cf-gh}), the algorithm
becomes in our particular case
\begin{eqnarray}
u_i(t+\Delta t)=u_i(t)+f_i\left(u(t)\right)\Delta t+ u_i(t) X_i
\nonumber \\
+ \frac{1}{2} u_i(t) X_i^2 +Y_i + \frac{1}{2} X_i Y_i
\label{eq:part-alg}
\end{eqnarray}
(now repeated indexes are {\em not} summed up). The last term in
this expression can be replaced by its statistical average (which
is zero here), since it is known that in this case the resulting
algorithm represents a stochastic process having the same
statistical properties \cite{ramirez93b}. Hence the
algorithm which we use is finally:
\begin{eqnarray}
u_i(t+\Delta t)=u_i(t)+f_i\left(u(t)\right)\Delta t+ u_i(t) X_i
\nonumber \\
\label{eq:shc-alg}
+ \frac{1}{2} u_i(t) X_i^2 + Y_i
\end{eqnarray}

The space-correlated random field $Z$ will be generated from the
following relation:
\begin{equation}
\label{eq:rf-gen-1}
Z\left(\vec{x}\right)\;=\;exp\left( \frac{1}{2\pi} \lambda^2
\nabla^2\right)\,W\left(\vec{x}\right)\,,
\end{equation}
which is written in continuum space. $W\left(\vec{x}\right)$ is
a gaussian white random field of intensity $D$:
\begin{equation}
\label{eq:rf-gen-2}
\left<W\left(\vec{x}\right)\;W\left(\vec{x}\,'\right)\right>\;=
\;D\,\delta\left(\vec{x}-\vec{x}\,'\right)
\end{equation}
$Z\left(\vec{x}\right)$ can be easily generated in Fourier space,
where $W$ becomes an anti-correlated field
\cite{ojalvo92b}. Definition (\ref{eq:rf-gen-1}) is chosen
in such a way that random field $Z\left(\vec{x}\right)$
(and hence multiplicative noise $\xi\left(\vec{x},t\right)$)
have a well-defined correlation length $\lambda$, as can
be seen by computing the space correlation function
$D\left(\frac{\vec{x}-\vec{x}\,'}{\lambda}\right)$, which
can be done in a straightforward way by writing
(\ref{eq:rf-gen-1}) in Fourier space. The result is:
\begin{equation}
\label{eq:rf-gen-3}
D\left(\frac{\vec{x}-\vec{x}\,'}{\lambda}\right)\;=\;
\frac{D}{4\lambda^2}\,exp\left( -\frac{\pi}{4}\,
\frac{\mid\vec{x}-\vec{x}\,'\mid^2}{\lambda^2}\right)
\end{equation}

\subsection{Analysis of the transition}

\begin{figure*}
\centerline{\psfig{figure=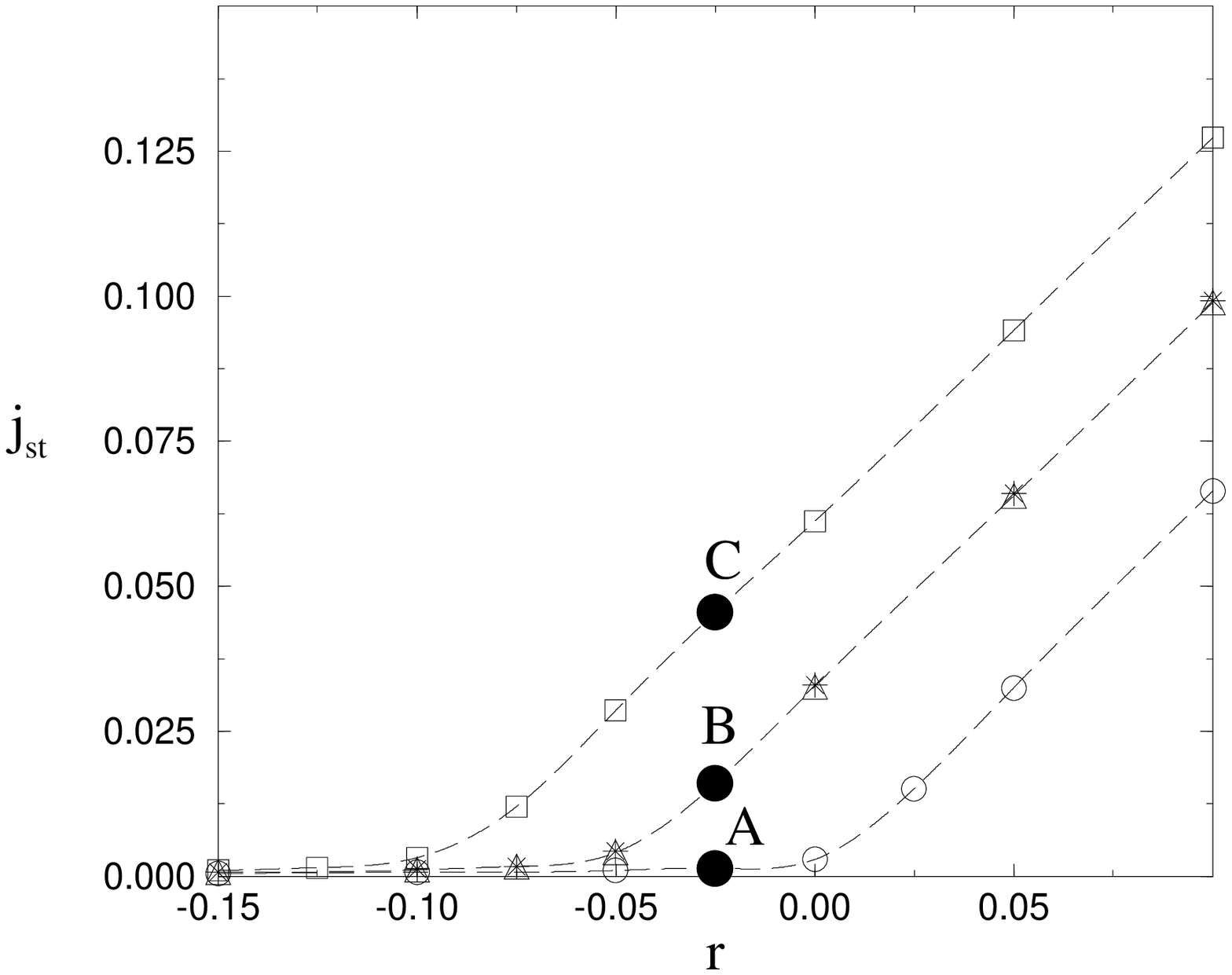,height=6cm}}
\caption{\em
Steady convective flux vs. control parameter $r$ in the absence
of external noise (circles), with a white multiplicative noise 
of intensity $D=0.1$ (squares) and with a multiplicative noise
of intensity $D=0.1$ and correlation length
$\lambda=0.71$
(triangles and stars).
System size is $40 \times 40$ except for the stars, which correspond
to $30 \times 30$.
\label{fig:bd}
}
\end{figure*}

In order to analyse the pattern-forming bifurcation exhibited by model
(\ref{eq:dgshe}-\ref{eq:cf-gh}), we shall define
the following quantity:
\begin{equation}
\label{eq:chf}
J(t)\;=\;\left< \int u^2\left(\vec{x},t\right)
\;d\vec{x}\right>\;,
\end{equation}
where the statistical average is made over realizations of both
the internal and external noises.
In the Rayleigh-B\'enard case,
the density of this quantity ($j(t)=J(t)/V$)
corresponds to the density of heat flux due to convection from
the lower towards the upper plate of the cell. Hence its value
is zero in the homogeneous state (no convective rolls) and non-zero
in the structured convective phase, increasing linearly with
the control parameter $r$, as shown clearly by experiments.
This behaviour is recovered by numerical simulations of the
model. In the absence of external noise, stripes appear for
values of $r$ greater than
$r\simeq 0$, as discussed earlier. This can be seen in Fig.
\ref{fig:bd}, where the steady-state value of the heat flux is
plotted against the control parameter of the system.
It should be noted that the existence of a small but non-zero
internal additive noise, along with the fact that simulations
are performed on small finite systems, produces a rounding-off
of the otherwise sharp transition described above.
Periodic boundary conditions are considered,
and $\Delta x=1$ is chosen. Given this lattice spacing,
a value of $\Delta t=0.01$ happens to be enough to
ensure stability of the algorithm.
The intensity of the additive noise is taken to be equal to
$10^{-3}$ in all cases. On the other hand, the value of
the wavenumber $k_0$ is chosen in each case so that each convective roll
is described by $10$ lattice cells.

In the presence of a non-zero external multiplicative uncorrelated
($\lambda=0$) noise the bifurcation is shifted to the conducting
($r<0$) region, which leads to the appearance of roll patterns in
otherwise subcritical regions (squares in Fig. \ref{fig:bd}). The amount
of the shift is decreased back by the introduction of space correlation
($\lambda\neq 0$) in the external fluctuations, as shown
by the empty-triangle curve in Fig. \ref{fig:bd}. In this last case the
simulation is also performed for a smaller value of the system
size, but no finite-size effects are encountered (asterisks
in the same figure).

\begin{figure*}
\centerline{\psfig{figure=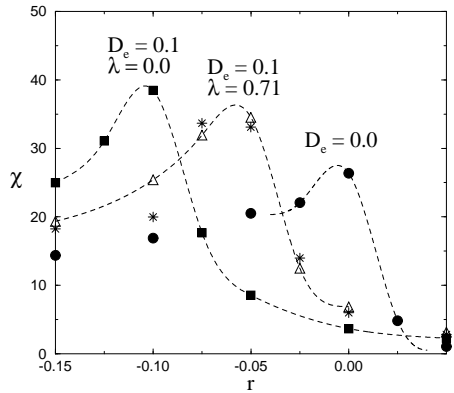,height=6cm}}
\caption{\em
Relative fluctuations of the steady convective heat flux vs.
control parameter $r$. Values of the parameters are the same
as those in Fig. 1.
\label{fig:sd}
}
\end{figure*}

In analogy to studies of equilibrium phase transitions, the
bifurcation shift found above can also be observed
from the computation of the relative fluctuations of the
steady heat flux, which are expected to exhibit some sort
of singular behaviour in the transition point. Indeed, if
one computes the relative fluctuations of $J$ as
\begin{equation}
\label{eq:susc}
\chi\;=\;V\;\frac{\left<J^2\right> - \left<J\right>^2}
{\left<J\right>^2}\;\;,
\end{equation}
then the transition point is characterised by a maximum value of
this quantity. This is observed in Fig. \ref{fig:sd}, where the transition
shifts, which were already present in Fig. \ref{fig:bd}, are clearly
observed. Again no finite-size effects are found.
On the other hand, unfortunately the analysis of neither $J$ nor
its relative fluctuations $\chi$ gives a hint on the effect of
multiplicative noise on the nature of the bifurcation (i.e. on
the order of the transition). Further analytical work would be
needed in order to clarify this point.

\begin{figure*}
\centerline{{\large (a)}\hskip1cm
\psfig{figure=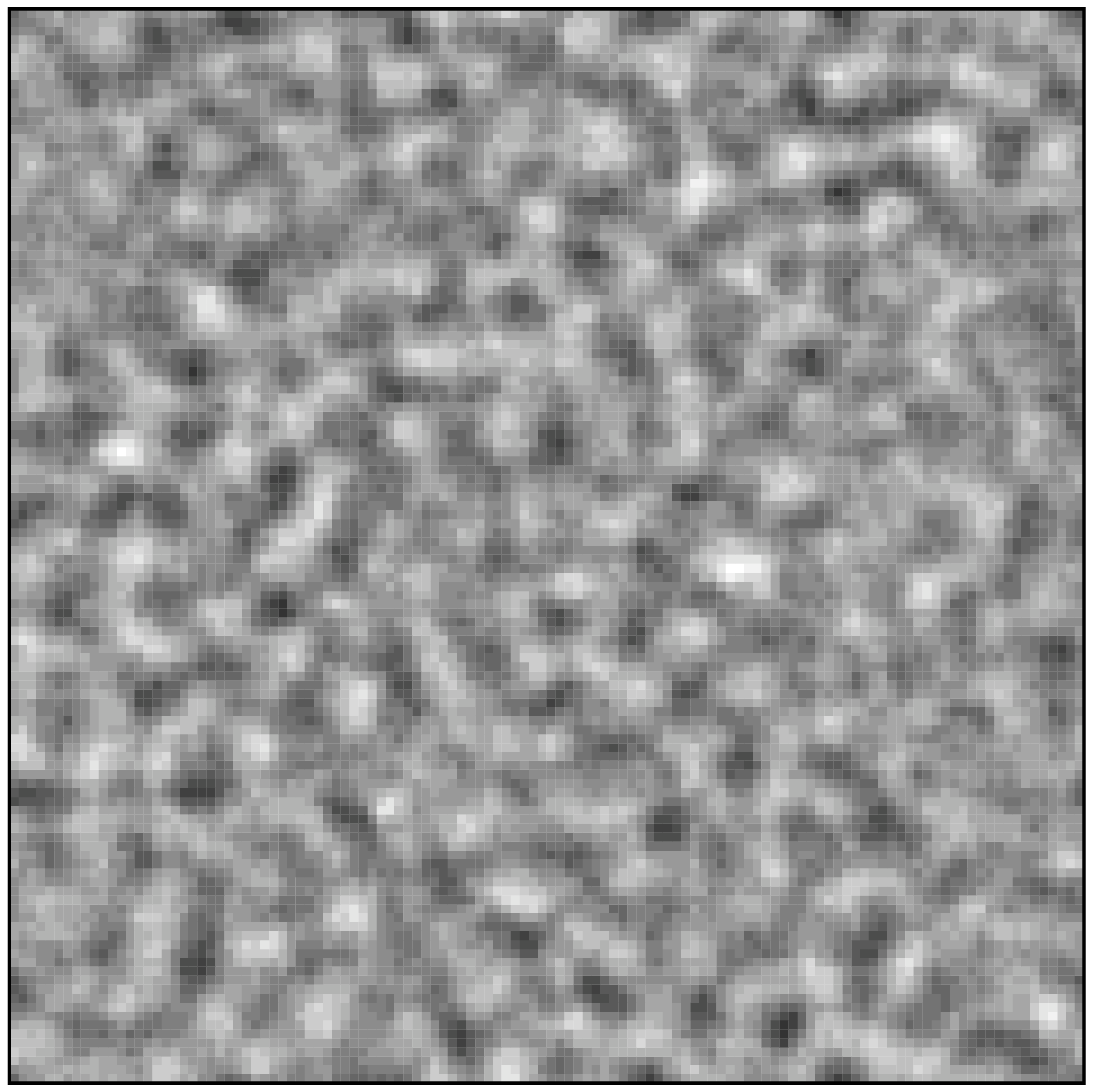,height=6cm}}
\centerline{{\large (b)}\hskip1cm
\psfig{figure=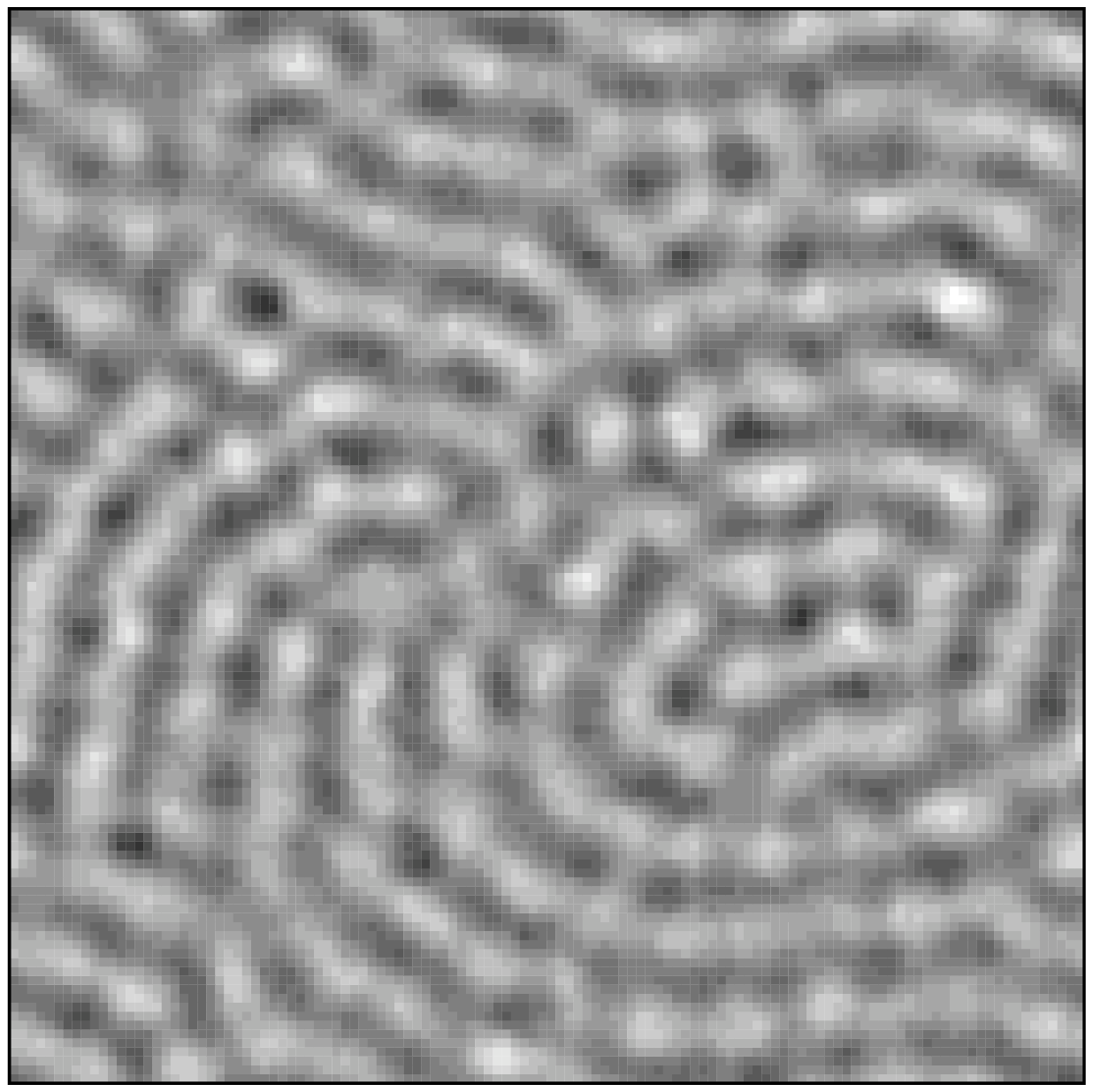,height=6cm}}
\centerline{{\large (c)}\hskip1cm
\psfig{figure=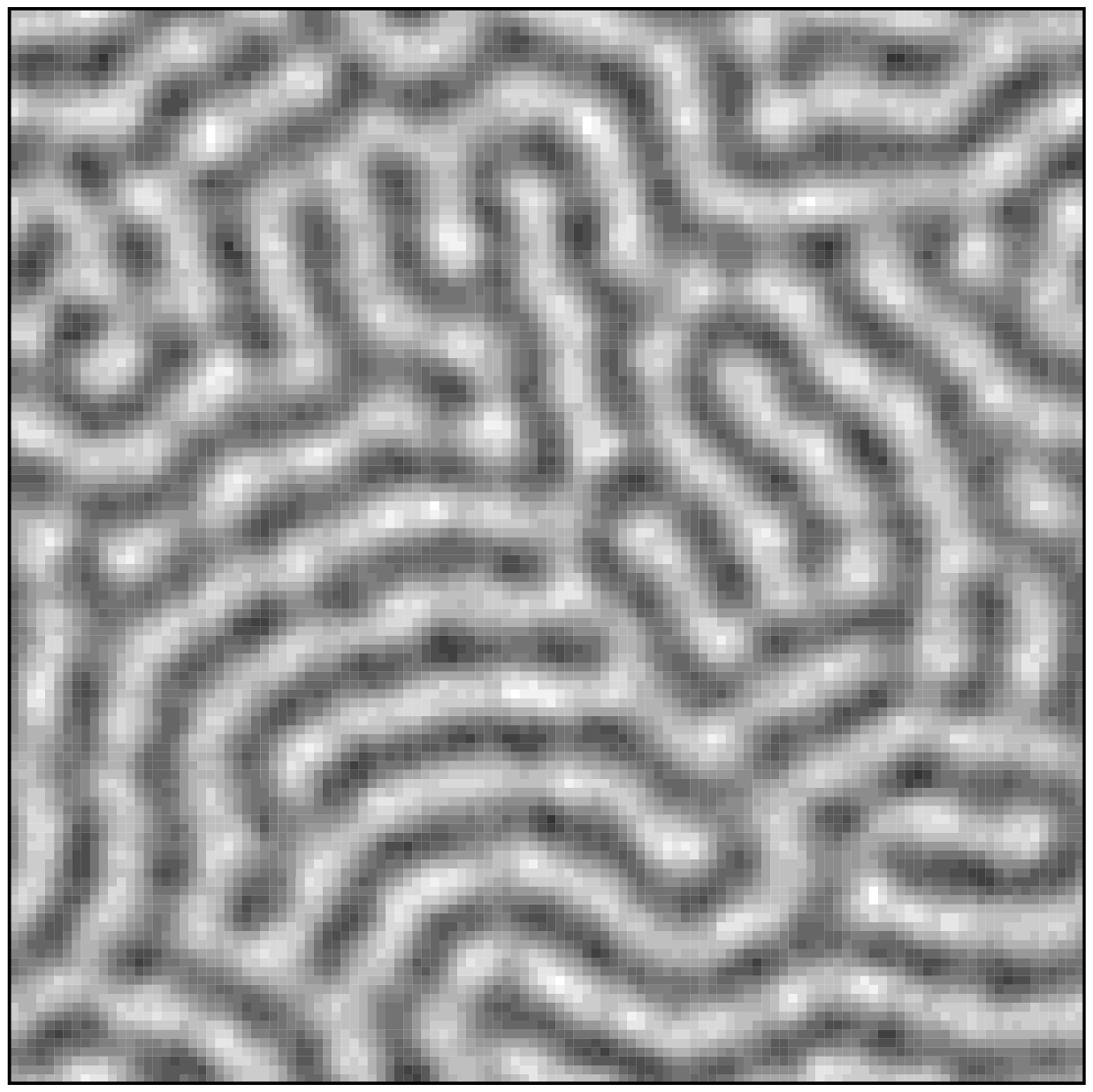,height=6cm}}
\vskip7mm
\caption{\em
Patterns corresponding to the three situations named $A$, $B$ and
$C$ in Fig. 1.
\label{fig:pat}
}
\end{figure*}

In conclusion, simulations show that stripe patterns can be favored
by multiplicative noise, this effect being diminished by correlation
of the noise. Figure \ref{fig:pat} shows three different
situations corresponding
to the states marked A, B and C in Fig. \ref{fig:bd}. Cases B and C are
patterns for $r<0$ and hence {\em favored} by multiplicative noise.

\subsection{Structure function in the convecting phase}

In order to analyse more deeply the lack of finite-size effects in the
convective heat flux above threshold, we have computed the stationary
spherically-averaged {\em structure function} in a noise-favored convecting
state (with a negative $r$ but a supercritical value of $D$).
This function is defined as the Fourier transform of the
correlation function of the system,
\begin{equation}
\label{eq:cf}
G\left(\vec{x},t\right)\;=\;\frac{1}{V}\,\int_V\,d\vec{x}\,'\left<
u\left(\vec{x}\,',t\right)\,u\left(\vec{x}\,'+\vec{x},t\right)\right>
\end{equation}

Hence, the structure function indicates
any periodicity in the system, so that it is a good way of characterising
a pattern. Let us consider a
finite system of volume $V$. The Fourier transform $u_\mu(t)$
of the field $u\left(\vec{x},t\right)$ is defined by:
\begin{equation}
\label{eq:ft}
u\left(\vec{x},t\right)\;=\;\frac{1}{V}\,\sum_\mu e^{i \vec{k}_\mu
\cdot \vec{x}}\,u_\mu
\end{equation}
where the components of $\vec{k}$ have the form $2\pi\mu/L$, where
$\mu$ is an integer and $L$ is the corresponding linear dimension
of the system. Then it can be seen that the structure function is:
\begin{equation}
\label{eq:sf}
S_\mu\;=\;\frac{1}{V}\,\left<u_{\mu}\,u_{-\mu}\right>
\end{equation}

\begin{figure*}
\centerline{\psfig{figure=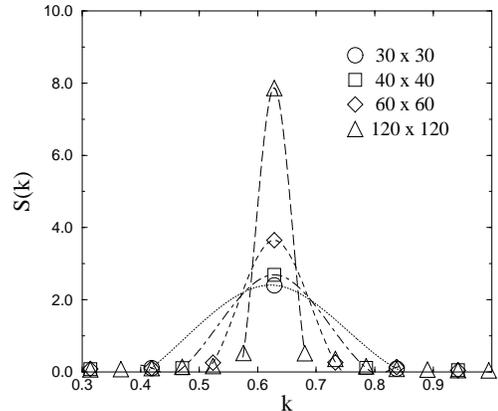,height=6cm}}
\caption{\em
Structure function for the same convecting state with different
system sizes.
\label{fig:sk-above}
}
\end{figure*}

\begin{figure*}
\centerline{\psfig{figure=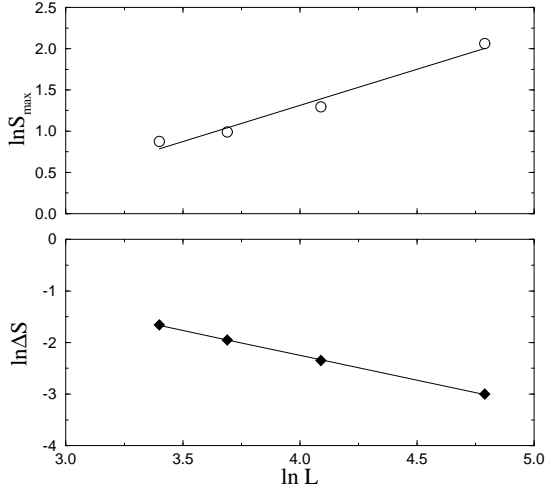,height=6cm}}
\vskip1cm
\caption{\em
Logarithmic plots of the maximum and width of the structure
function against system size. Linear fitting gives exponents
$\sim+1$ and $\sim-1$, respectively.
\label{fig:sk-mw}
}
\end{figure*}

Simulations show that this function behaves in the same way
for the whole parameter range, presenting a unique maximum
related to the periodicity of the pattern in the convecting
phase and to thermal fluctuations in the conducting one,
even though in this last case the height of the peak is
several orders of magnitude lower than in the convecting
one. Calculations have been performed for different system
sizes in the patterned region, where the peaks are all seen to
be centered at approximately the same value of $k$, $k\sim k_0$ (in this case
$k_0\simeq 0.63$), but their heights increase and their widths decrease
with increasing $L$ (see Fig. \ref{fig:sk-above}).
In fact, it can be seen (Fig. \ref{fig:sk-mw})
that $S_{max}\sim L$ and $\Delta S\sim L^{-1}$, so that the total
area under the curve $S(k)$ is kept constant with varying $L$. This result
agrees with the lack of finite-size effects in $J_{st}$, since
the convective heat flux can be seen to be equal to the area under the
structure function. All these
facts lead to the conclusion that the structure function approaches
a delta function in the thermodynamic limit (where the rolls are
perfectly shaped) as the system size increases.

\section{Linear stability analysis of the structure function}
\label{sec:sth}
A linear stability analysis of the homogeneous $u=0$ solution of model
(\ref{eq:msh}) permits to estimate the transition point as that
at which this solution becomes unstable. Beyond it the system
departs from zero and becomes saturated by the nonlinearity, giving
rise to the appearance of a non-zero roll solution.
Due to the relevant role of the structure function as a means of
characterising a patterned state, as has been discussed in the
previous section, we shall begin our theoretical approach to
the problem with a linear stability analysis of this function.

We now assume that the homogeneous conducting $u=0$ solution
is affected by small perturbations, so that we only need to
evaluate the evolution equation of $S_\mu$, defined in (\ref{eq:sf})
in the {\em linear} regime. This can be done
by using the Fokker-Planck equation governing the behaviour in
time of the probability density of the field (in this case in
Fourier space). As shown in the Appendix,
the Fokker-Planck equation for
the probability distribution of the stochastic process
$u_\mu(t)$ is
\begin{eqnarray}
\frac{\partial P}{\partial t}\;=\;- \sum_\mu \frac{\partial}
{\partial u_\mu} F_\mu P \; +\; V\varepsilon\;
\sum_{\mu,\nu} \; \frac{\partial}{\partial u_\mu}
\;\frac{\partial}
{\partial u_{-\mu}} \; P \nonumber \\
+\; \frac{1}{V}\;
\sum_{\mu,\nu,\nu'}\; \frac{\partial}{\partial u_\mu}
\;G_{\mu-\nu'}\;D_{\nu'}\;\frac{\partial}
{\partial u_\nu} \;G_{\nu+\nu'}\; P
\label{eq:fpe2-6}
\end{eqnarray}
where $D_\nu$ is the discrete Fourier transform of $D_{i-j}$, and
$F$ and $G$ are the Fourier transforms of the
corresponding terms in the discretized version of the SH equation
(\ref{eq:msh}):
\begin{equation}
\label{eq:dmsh}
\frac{d u_i}{d t}\;=\;f_i\left(\vec{u}(t)\right)\;+\;
\eta_i(t)\;+\;g_i\left( \vec{u}(t)\right)\;
\xi_i(t)
\end{equation}
So that
\begin{eqnarray}
\label{eq:efandgi}
f_i=\left(r-\left(\nabla_{ij}^2+k_0^2\right)^2\right)u_j
\;\;\;\;\;\;\;\;\;\;\;\;\;\;\;\;\;\;\;\;\;\;\;\;\;\;\;\;\;
\nonumber \\
\leftrightarrow\;
F_\mu=\left(r-\left(\nabla_\mu^2+k_0^2\right)^2\right)
u_\mu \nonumber \\
g_i=u_i \leftrightarrow\; G_\mu=u_\mu
\;\;\;\;\;\;\;\;\;\;\;\;\;\;\;\;\;\;\;\;\;\;\;\;\;\;\;\;\;
\end{eqnarray}
where $\nabla_{ij}$ and $\nabla_\mu$ are a discretized version of
the Laplacian and its Fourier transform, respectively.
The time evolution equation for $S_\mu$ is given by:
\begin{equation}
\label{eq:sfeqfp1}
\frac{dS_\mu}{dt}=\frac{1}{V}\;\int {\cal D}u\; u_\mu u_{-\mu}\;
\frac{\partial P}{\partial t}
\end{equation}

\begin{figure*}
\centerline{\psfig{figure=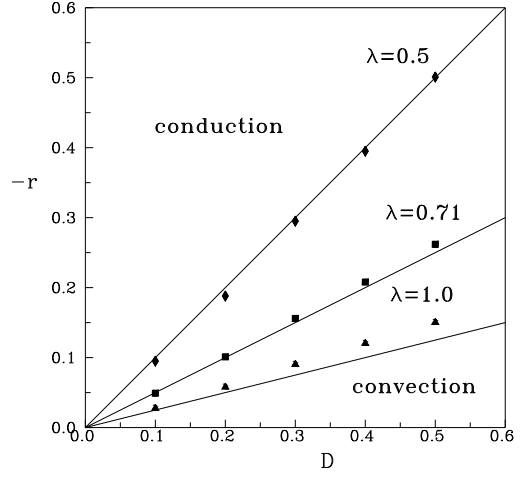,height=6cm}}
\vskip1cm
\caption{\em
Phase diagram of the system in the ($D, -r$) plane for
different values of the correlation length of the noise.
\label{fig:pd}
}
\end{figure*}

When Fokker-Planck equation (\ref{eq:fpe2-6}) is introduced in
this expression and integration by parts is performed, the
following equation is obtained:
\begin{eqnarray}
\label{eq:sfeqfp2}
V\;\frac{dS_\mu}{dt}= \left< u_\mu F_{-\mu} \right> +
\left< u_{-\mu}F_\mu\right>
\;\;\;\;\;\;\;\;\;\;\;\;\;\;\;\;\;
\nonumber \\
+ \frac{1}{V}
\int {\cal D}u \sum_{\eta,\nu,\nu'}
\delta_{\mu \eta}\;\delta_{-\mu
\nu}\;G_{\eta-\nu'}\;D_{\nu'}\;G_{\nu+\nu'}\;P
\nonumber \\
+ \frac{1}{V} \int {\cal D}u
\sum_{\eta,\nu,\nu'}
\delta_{\mu \eta}\;u_{-\mu}\;\frac{\partial
G_{\eta-\nu'}}{\partial
u_\nu}\;D_{\nu'}\;G_{\nu+\nu'}\;P
\nonumber \\
+ \frac{1}{V}\int {\cal D}u
\sum_{\eta,\nu,\nu'}
\delta_{-\mu \eta}\;\delta_{\mu
\nu}\;G_{\eta-\nu'}\;D_{\nu'}\;G_{\nu+\nu'}\;P
\nonumber \\
+ \frac{1}{V} \int {\cal D}u
\sum_{\eta,\nu,\nu'}
\delta_{-\mu \eta}\;u_{\mu}\;\frac{\partial G_{\eta-\nu'}}
{\partial u_\nu}\;D_{\nu'}\;G_{\nu+\nu'}\;P\;
\nonumber \\
+\;2\,\varepsilon\;V
\end{eqnarray}
Introduction of (\ref{eq:efandgi}) into
this expression leads to
\begin{eqnarray}
\frac{d}{dt}S_\mu(t)\; =\; 2 \left(r-\left(
\nabla_\mu ^2 +k_0^2\right)^2\right) S_\mu(t) + 2 \varepsilon +
\nonumber \\
\frac{2}{V}\; S_\mu \;\sum_\nu D_\nu
+\;\frac{2}{V^2} \;\sum_\nu D_\nu
\left<u_{\mu-\nu}\,u_{-\mu+\nu}\right>
\label{eq:sfeqfp9}
\end{eqnarray}
But, according to definition (\ref{eq:ft}), the
following relation holds:
\begin{equation}
\label{eq:d0}
\sum_\nu D_\nu =  V \; D(\vec{x}=0)  = V\; D(0)
\end{equation}
Thus the equation for the structure function is finally
\begin{eqnarray}
\frac{d}{dt}S_\mu(t)\; =\;
2 \left(r + D(0) -\left(\nabla_\mu ^2+k_0^2\right)^2 \right)
S_\mu(t)
\nonumber \\
+ 2 \varepsilon +
2 \;\frac{1}{V}\; \sum_\nu D_\nu S_{\mu-\nu}
\label{eq:sfeqfp8}
\end{eqnarray}
Translation of this equation into its continuum version leads to
the final expression for the evolution equation of the structure
function of the Swift-Hohenberg model in the presence of a
space-colored multiplicative external noise:
\begin{eqnarray}
\label{eq:feesf}
\frac{\partial}{\partial t}S(\vec{k},t) = 2 \left[r +
D(0) - \left( k^2-k_0^2 \right)^2 \right] S(\vec{k},t)
\nonumber \\
+2 \varepsilon + 2 \frac{1}{(2\pi)^2}
\int D(\vec{q}) S(\vec{k}-\vec{q},t) d \vec{q}
\end{eqnarray}
The stability analysis of this equation needs some comments. The last
term in Eq. (\ref{eq:feesf}) is a mode-coupling term, which is essentially
non-linear and will be discarded in our linear analysis.
In this case, the presence of a multiplicative noise leads to the
existence of an effective noise-dependent control parameter, $r + D(0)$,
so that any perturbation of the homogeneous state will always grow if
the condition $r+D(0)>0$ is obeyed. Hence, linear analysis predicts
that, under the presence of multiplicative noise, the system can leave
the homogeneous state in situations for which $r<0$. The
nonlinearity will then stabilise the system in an ordered state.
Besides, $D(0)=D/(4\lambda^2)$ for the particular noise spectrum
we have chosen, so that the amount of the shift decreases with
increasing correlation length. This effect can be observed in the
simulation results presented in Fig. \ref{fig:pd}, where a phase diagram
of the system is plotted in the $(D,-r)$ plane for different values
of the correlation length of the noise $\lambda$. It is worth noting
that, as long as $D \neq 0$, the critical value of $r$ for the transition
from conduction to convection is strictly negative. As predicted by
our analysis, the critical curve in this plane is a straight line whose
slope decreases as $\lambda$ increases, reducing the noise-favored region
and thus the shift effect due to the multiplicative noise. In the figure,
isolated symbols are transition points as obtained from
simulations, whereas solid lines are the corresponding
relations coming from the linear analysis ($r+D/(4\lambda^2)=0$).

In the linear stability analysis of Ref. \cite{becker94}, the
last term of Eq. (\ref{eq:feesf}) was considered and was proved to give
corrections of order $D(0)^2$. On the other hand, the relevance of
this term can be seen by studying the stationary structure function.
From evolution equation (\ref{eq:feesf}),
this function can be analytically found in the subcritical case
($r+\frac{D}{\Delta x^2} < 0$) for a white external noise
($D(\vec{k})=D$), the result being:
\begin{equation}
\label{eq:sf-th}
S_{st}(k)\,=\,\frac{\varepsilon'}{\left(k^2-k_0^2\right)^2+ \mid
r_{eff}\mid}\,=\,\frac{\varepsilon'}{\omega(k)}
\end{equation}
where the effective control parameter $r_{eff}$ is
\begin{equation}
\label{eq:eff-cp}
r_{eff}\,=\,r + \frac{D}{\Delta x^2}
\end{equation}
and the renormalised additive-noise intensity is
\begin{equation}
\label{eq:ren-ani}
\varepsilon'\,=\,\frac{\varepsilon}{1-\gamma D}
\end{equation}
with
\begin{eqnarray}
\gamma=-\frac{1}{4\pi}\,\mid r_{eff}\mid^{-1/2}\,arctg \mid
r_{eff} \mid^{1/2 }
\nonumber \\
= \frac{1}{(2\pi)^2}\,\int\frac{1}{\omega(k)}
\,d\vec{k}
\label{eq:gamma-par}
\end{eqnarray}

\begin{figure*}
\centerline{\psfig{figure=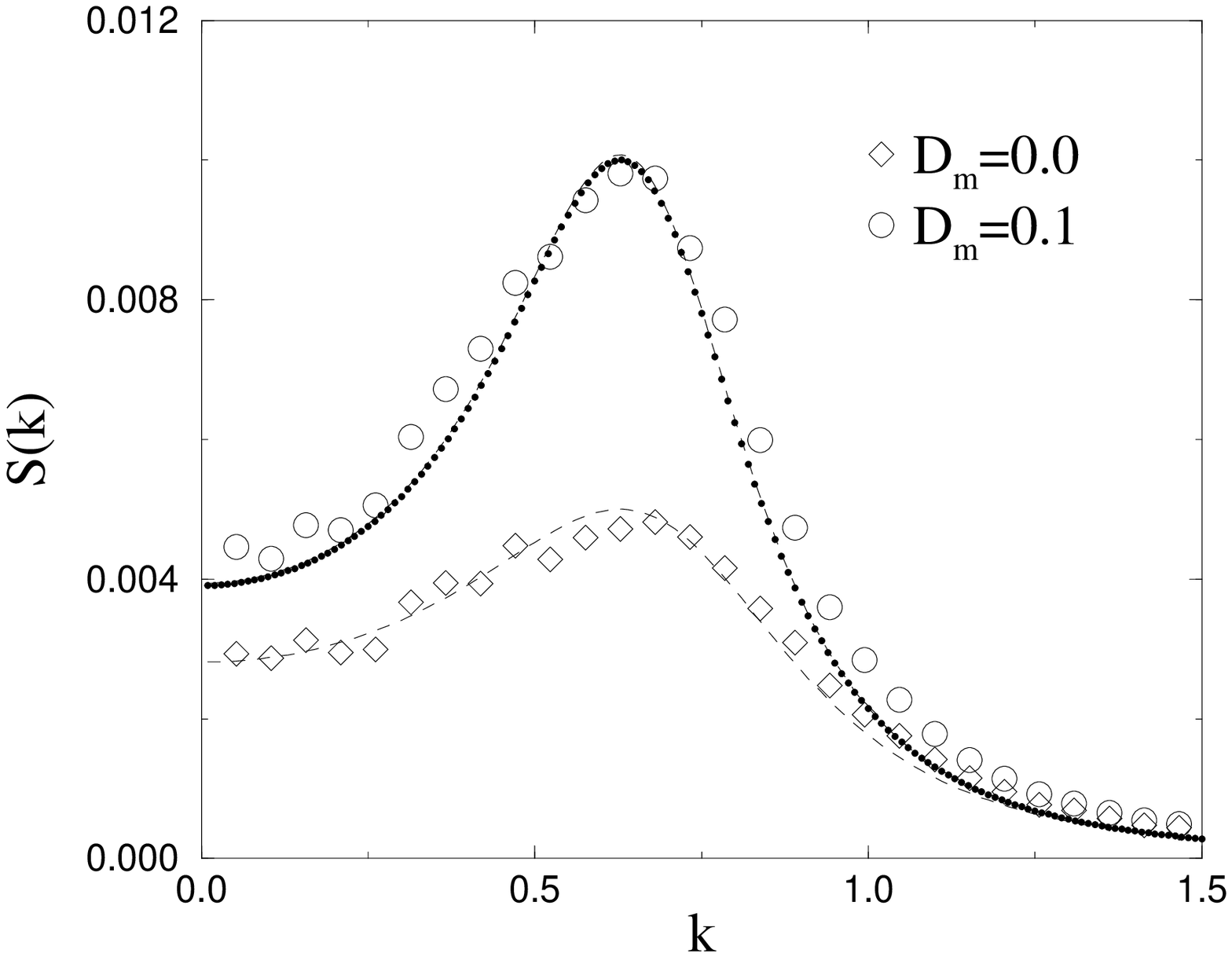,height=6cm}}
\caption{\em
Simulation results for the spherically-averaged structure function
for a conducting state with (circles) and without (diamonds)
multiplicative noise. Dashed lines are the continuum analytical
solutions coming from linear analysis, and the dotted line corresponds
to the solution with no mode-coupling term ($\gamma$ set equal to $0$).
Values of the parameters are $r=-0.2$, $\varepsilon=0.001$,
$\Delta x = 1.0$ and $k_0=0.63$.
\label{fig:sk-below}
}

\centerline{\psfig{figure=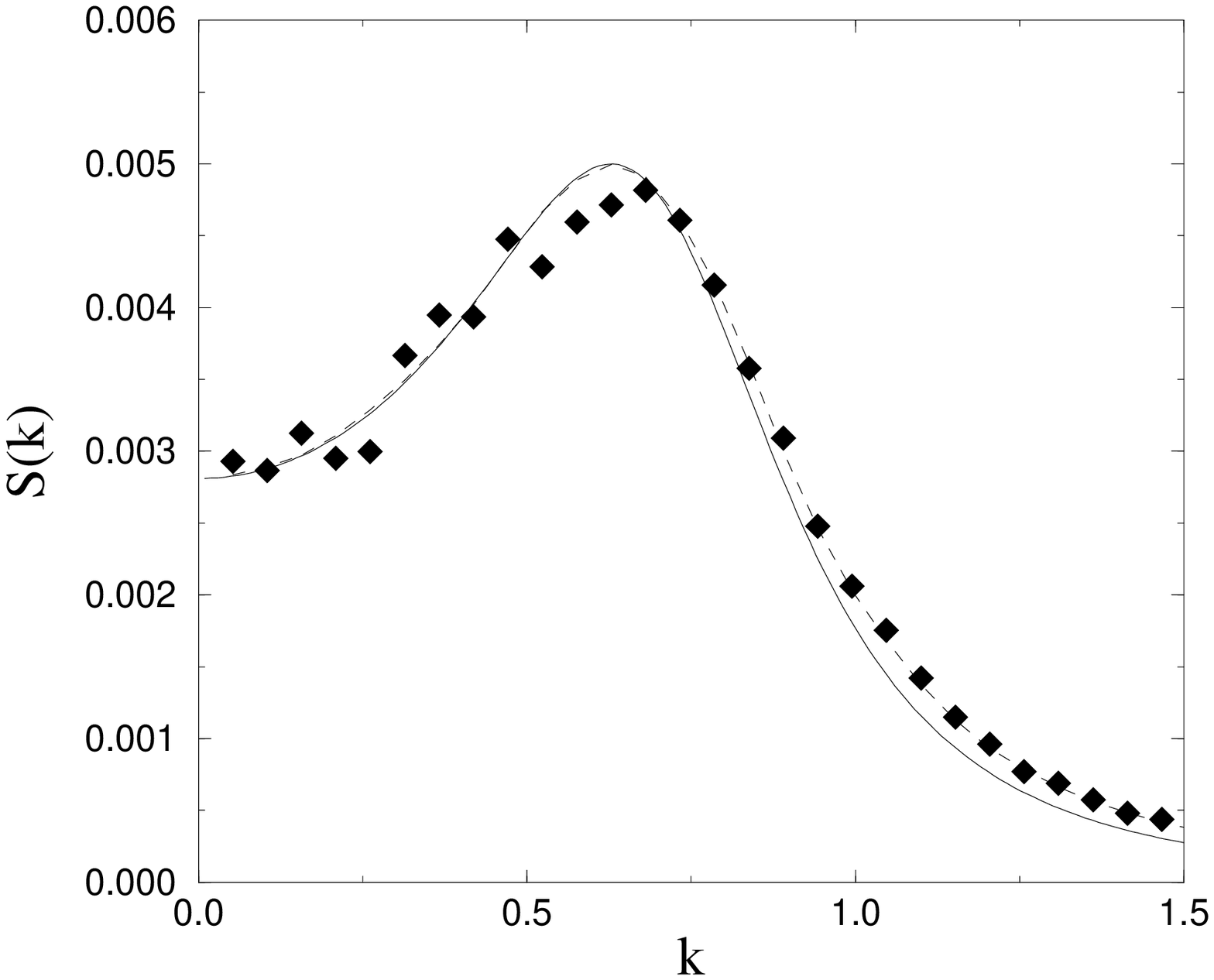,height=6cm}}
\caption{\em
Comparison of the spherically-averaged structure function obtained
by simulation (diamonds), discrete integration of the analytical
linear evolution equation (dashed line) and linear
analytical continuum solution (solid line).
\label{fig:lin-disc-comp}
}
\end{figure*}

This linear result can be compared satisfactorily with (nonlinear)
numerical simulations, since in the subcritical region nonlinear
terms are expected to be negligible. Figure \ref{fig:sk-below}
shows both kind of results with and without multiplicative noise.
The ordering role of this external noise is observed even in this
conducting situation, enhancing the maximum value of the structure
function at $k=k_0$. The growing discrepancy between theory and
simulation at large wavenumbers can be understood from the fact
that we are comparing an analytical result derived in a continuum
space with numerical results coming from a discrete-space
simulation. Figure \ref{fig:lin-disc-comp} shows a comparison (for
the case $D=0$) between these two results and a (discrete) numerical
integration of Eq. (\ref{eq:feesf}). The agreement between these
results and the simulation ones is quite good even at large wavenumbers.

It should be noted that making $\gamma=0$ in
(\ref{eq:sf-th})-(\ref{eq:gamma-par}) amounts to ignoring the
mode-coupling term in Eq. (\ref{eq:feesf}). Figure \ref{fig:sk-below}
shows that the result (\ref{eq:sf-th}) with $\gamma=0$ is almost
identical to the solution of the full linear analysis, in agreement
with our previous statement concerning the lack of influence of
the mode-coupling term in a first-order linear stability analysis.

\section{Linear stability analysis of n-th order moments}
\label{sec:th}
Due to the stochastic character of the field $u\left(\vec{x},t\right)$,
the linear stability analysis can also be done on the statistical
moments of $u$. In this case, differences between the zero-dimensional
and the spatially-extended cases appear, as will be discussed in
what follows.

\subsubsection{Zero-dimensional case}

Let us consider the following zero-dimensional model, known as
the Stratonovich model:
\begin{eqnarray}
\frac{d x}{d t}\;=\;\alpha\,x - x^3 + x\,\xi(t)
\label{eq:strat}
\\ \left< \xi(t)\,\xi(t') \right> = 2\,D\,\delta(t-t')\nonumber
\end{eqnarray}

In the absence of noise, this model exhibits a supercritical pitchfork
bifurcation at $\alpha=0$. For negative values of $\alpha$, the stable
stationary solution of (\ref{eq:strat}) is $x=0$. In the presence of
the multiplicative noise, position of the bifurcation point can be given
by a linear stability analysis of solution $x=0$. Since $x$ is now a
stochastic process, the stability analysis is performed on its statistical
moments, whose time evolution can be found from the Fokker-Planck equation
governing the evolution of $P$. In the linear regime we have,  
\begin{equation}
\label{eq:lsa-0d-2}
\frac{\partial P}{\partial t}\;=\;-\frac{\partial}{\partial x}
\,\alpha x\, P\;+\;D\,\frac{\partial}{\partial x}\,x\,
\frac{\partial}{\partial x}\,x\,P
\end{equation}
where $P(x,t)$ is the probability density of the stochastic process
$x(t)$. This equation leads to
\begin{equation}
\label{eq:lsa-0d-3}
\frac{d}{d t}\left<x^n\right>\;=\;n\,(\alpha + n\,D)\,\left<x^n\right>
\end{equation}
which indicates that the bifurcation point for the n-th order moment
is located at $\alpha_c=-n\,D$. Hence the position of the bifurcation
point depends on the order of the statistical moment which is being analysed,
which makes this analysis meaningless. We will see shortly that spatial
coupling in extended systems introduces changes to this situation,
leading to a useful analytical result.

\subsubsection{Spatially-extended case}

We now turn our attention back to the SH model with multiplicative
noise (\ref{eq:msh}). In order to obtain a possible higher-order
generalisation of the structure function, let us consider the following
correlation function:
\begin{equation}
\label{eq:gcf}
G^{(n)}\left(\vec{x},t\right)\;=\;\frac{1}{V}\,\int_V\,d\vec{x}\,'\left<
\prod_{l=1}^{n}\,u\left(\vec{x}\,'+(l-1)\vec{x},t\right)\right>
\end{equation}
which measures the correlation between the values of the field
at $n$ points equally spaced along a line. Its Fourier transform
leads to a generalised n-th order structure function, 
\begin{eqnarray}
S^{(n)}_\mu\;=\;\frac{1}{V}\,\sum_{\mu_1}\,\sum_{\mu_2}\cdots\sum_{\mu_n}
\,\left<\prod_{i=1}^{n}\,u_{\mu_i}
\right>
\nonumber \\
\times \delta_{\sum_{i=1}^{n}\mu_i,0}\,\delta_{\mu+
\sum_{i=1}^{n}(i-1)\mu_i,0}
\label{eq:gsf}
\end{eqnarray}

Applying to this quantity the procedure described in the previous
paragraphs, making use of the Fokker-Planck equation (\ref{eq:fpe2-6}),
performing integration by parts and ignoring internal noise, one
finds
\begin{eqnarray}
\frac{d}{dt}S^{(n)}_\mu(t)\; =\;
n \left(r + D(0) -\left(\nabla_\mu ^2+k_0^2\right)^2 \right)
S^{(n)}_\mu(t)
\nonumber \\
+\,\frac{n(n-1)}{V}\; \sum_\nu \,
D_\nu \,S^{(n)}_{\mu-\nu}
\label{eq:gsfeqfp8}
\end{eqnarray}

Unlike the zero-dimensional case, all dependence on $n$ of the
linear coefficient in the previous equation is factorised out.
Hence, this results shows that spatial coupling prevents the linear
analysis from giving different first-order results
for different-order statistical
moments (part of the contribution of the multiplicative noise
comes through the inhomogeneous term in Eq. (\ref{eq:gsfeqfp8}),
which represents the coupling between spatial modes).
There exist, of course, other possible methods to study and extract
information from the behaviour of higher-order moments in
pattern-forming systems (see, for instance \cite{mikhailov92}).

\section{Conclusion}
Simulations of the Swift-Hohenberg model with a space-correlated
fluctuating control parameter show that the bifurcation point
from a homogeneous to a structured state is shifted from the
standard additive-noise case. The amount of the shift increases
with increasing intensity of the multiplicative noise, whereas
space correlation in the fluctuations shifts back the transition
point towards the standard situation. A good qualitative and
quantitative agreement is obtained between this numerical study
and a linear stability analysis of the structure function of the
system, for the values of the parameters used here. The stability
analysis has also been performed on a set of generalised structure
functions which correspond to higher-order statistical moments,
and the same results are found in all cases (i.e. for all
orders) within our assumptions (i.e. at first order), which constitutes
a qualitative difference to the role of multiplicative noise in
zero-dimensional systems.

In conclusion, in this pattern-forming spatially extended system,
multiplicative noise produces an unambiguous shift of the transition
in the ordering direction. It is worth saying, however, that
according to recent studies on other extended systems in the
presence of multiplicative noise \cite{broeck94}, one can expect
that for large values of the intensity of this noise, the system
will return to the disordered state.

\begin{acknowledgements}
This research was supported in part by the Direcci\'on General de
Investigaci\'on Cient\'{\i}fica y T\'ecnica (Spain) under Project
No. PB93-0769. Most of the simulations were performed on the
CRAY Y-MP of the Centre de Supercomputaci\'o de Catalunya
(CESCA). Fruitful discussions with Prof. L. Kramer and J. Casademunt
are acknowledged.
\end{acknowledgements}

\appendix
\section*{Fokker-Planck equation for a spatially-extended process
with multiplicative noise}

In the following we shall assume a multiplicative noise with a
non-delta correlation only in space:
\begin{equation}
\label{eq:gencor}
\left< \;\xi(\vec{x},t) \;\xi(\vec{x}\,',t') \;\right> =
2\;D(\vec{x}-\vec{x}\,')\;\delta(t-t')
\end{equation}

Our aim now is to find the Fokker-Planck equation governing the
stochastic
process whose evolution is given by the general equation
\begin{equation}
\label{eq:langeq}
\frac{\partial u\left(\vec{x},t\right)}{\partial t}=
f\left(u\left(\vec{x},t\right)
,\nabla\right) + g\left(u\left(\vec{x},t\right) ,\nabla\right)
\xi\left(\vec{x},t\right).
\end{equation}
This can be
easily done in Fourier space. In a first step the equation is
written in a
discrete space, where the Langevin equation is:
\begin{equation}
\label{eq:dle}
\frac{d u_i}{d t} = f_i\left(u(t)\right)+
g_i\left(u(t)\right)\xi_i(t)
\end{equation}
where the cells have been named with one index independently of
the dimension of the discrete space. The correlation of the noise
in this discrete space is:
\begin{equation}
\label{eq:discor}
\left<\; \xi_i(t)\; \xi_j(t') \;\right> = 2\;
D_{i-j}\;\delta(t-t')
\end{equation}
where $D_{i-j}$ is the discrete version of the continuous
function $D$ describing the space decay of the correlation. In
the limit of zero correlation length (white-noise limit) it
becomes $\varepsilon\delta_{ij}/(\Delta x)^d$, where $\Delta x$
is the spacing of the lattice, i.e. the cell size.

Now we define the discrete Fourier transform and antitransform as
\begin{eqnarray}
\label{eq:pmu-pi}
u_\mu = \left(\Delta x\right)^{d}
\sum_i e^{-iq_\mu x_i} u_i \nonumber \\
u_i = \frac{1}{(L \Delta x)^{d}} \sum_\mu e^{iq_\mu x_i}
u_\mu
\end{eqnarray}
where $L$ is the number of cells per dimension,
$q_\mu = \frac{2\pi}{L\Delta x} \mu$, $x_i = i \Delta x$
and the sums are d-fold going from 1 to L. It can be seen that
the following relations hold:
\begin{eqnarray}
\label{eq:delta}
\sum_i e^{i(q_\mu-q_{\mu'})x_i} \;=\; L^d \;\delta_{\mu \mu'}
\nonumber \\
\sum_\mu e^{iq_\mu (x_i-x_j)}\; =\; L^d\; \delta_{ij}
\end{eqnarray}

The correlation of the noise in Fourier space, $\xi_\mu(t)$, can be
seen
to be
\begin{equation}
\label{eq:cnfs}
\left< \;\xi_\mu(t)\; \xi_\nu(t')\; \right> =
(L\Delta x)^{d}\;2\; D_\mu\;\delta_{-\mu,\nu}\;\delta(t-t')
\end{equation}
Here definitions (\ref{eq:pmu-pi}) and properties
(\ref{eq:delta})
have been used.

The equation of evolution of $u_\mu$ comes from the Fourier
transformation of Eq. (\ref{eq:dle}):
\begin{equation}
\label{eq:fdle}
\frac{d u_\mu}{d t}=f_\mu+\left(\Delta x\right)^{d}
\;\sum_i\;e^{-iq_\mu x_i} \;g_i\; \xi_i
\end{equation}
The last term at the right-hand side can be evaluated in terms
of the Fourier variables by means of (\ref{eq:pmu-pi}) and
(\ref{eq:delta}):
\begin{equation}
\label{eq:fdlebis}
\frac{d u_\mu}{d t}=f_\mu+ \frac{1}{(L\Delta x)^{d}}
\;\sum_{\nu}\;g_{\mu-\nu}\;\xi_\nu
\end{equation}
In the phase space of these Fourier variables we consider an
ensemble of systems corresponding to a given realization of the
noise and different initial conditions. The density of this
ensemble $\rho\left(u,t\right)$ must verify a continuity
Liouville equation:
\begin{equation}
\label{eq:liouv}
\frac{\partial\rho}{\partial t}\;=\;- \sum_\mu\;\frac{\partial}
{\partial u_\mu}\; \dot{u}_\mu \;\rho
\end{equation}
where $\rho\left(u,t\right)=\left<\delta
\left( u(t) -u \right) \right>_{IC}$,
the average being taken over initial conditions only. On the
other hand, the average of this density over the noise is the
probability density of the stochastic process
$P\left(u,t\right)= \left<\rho\right>$ (Van Kampen's
lemma \cite{vankampen76}) whose evolution equation is
the Fokker-Planck equation we
are looking for. Performing this noise average on
(\ref{eq:liouv}) leads thus to
\begin{equation}
\label{eq:fpemul1}
\frac{\partial P}{\partial t}= -\sum_\mu \frac{\partial}
{\partial u_\mu} f_\mu P - \frac{1}{(L\Delta x)^{d}}
\;\sum_{\mu,\nu} \;\frac{\partial}{\partial u_\mu}
\;g_{\mu-\nu}\; \left<\xi_\nu\rho\right>
\end{equation}
The average in the last term of the second member of this
equation can be evaluated by means of {\em Novikov's theorem}
\cite{novikov65},
which states that for a gaussian stochastic process $\xi$ the
following relation holds:
\begin{equation}
\label{eq:novi}
\left< \rho\left(u,t\right)\xi_\nu(t)\right>=
\int_0^tdt'\sum_\eta\left<\xi_\nu(t)\xi_\eta(t')\right>
\left<\frac{\partial \rho}{\partial\xi_\eta(t')}\right>
\end{equation}
and using (\ref{eq:cnfs}) one finds
\begin{equation}
\label{eq:novi2}
\left< \rho\left(u,t\right)\xi_\nu(t)\right>=
\left(L\Delta x\right)^{d} D_\nu\left<\frac{\partial \rho}
{\partial\xi_{-\nu}(t)}\right>
\end{equation}
And this last average can be calculated in the following way:
\begin{eqnarray}
\label{eq:aver}
\left<\frac{\partial \rho}{\partial\xi_{-\nu}(t)}\right>=
\;\;\;\;\;\;\;\;\;\;\;\;\;\;\;\;\;\;\;\;\;\;\;\;\;\;
\;\;\;\;\;\;\;\;\;\;\;\;\;\;\;\;\;\;\;\;\;\;\;\;
\nonumber \\
\sum_\eta\left<\frac{\delta u_\eta(t)}
{\delta\xi_{-\nu}(t)}\;\frac{\partial}
{\partial u_\eta(t)}\;\delta \left( u(t) -u
\right)\right>
\nonumber \\
=-\sum_\eta \frac{\partial}{\partial u_\eta}
\left<\left.\frac{\delta u_\eta(t)}{\delta\xi_{-\nu}(t)}
\right|_{u(t)=u}
\delta \left( u(t) -u\right)\right>
\end{eqnarray}
The functional derivative in this expression can be calculated
from Eq. (\ref{eq:fdlebis}). The result is:
\begin{equation}
\label{eq:aver2}
\frac{\delta u_\eta(t)}{\delta\xi_{-\nu}(t)}
=\frac{1}{\left(L\Delta x\right)^{d}}\,g_{\eta+\nu}
\left(u(t)\right)
\end{equation}
so that the Fokker-Planck equation we are looking for is finally
\begin{eqnarray}
\label{eq:fpemul2}
\frac{\partial P}{\partial t}= -\sum_\mu \frac{\partial}
{\partial u_\mu} f_\mu P
\;\;\;\;\;\;\;\;\;\;\;\;\;\;\;\;\;\;\;\;\;\;\;\;\;\;
\;\;\;\;\;\;\;\;\;\;\;\;\;\;
\nonumber \\
+ \frac{1}{(L\Delta x)^{d}}
\;\sum_{\mu,\nu,\eta} \;\frac{\partial}{\partial u_\mu}
\;g_{\mu-\nu}\; D_\nu\;\frac{\partial}{\partial u_\eta}
\;g_{\eta+\nu}\;P
\end{eqnarray}
In the particular case of a noise which is also white in
space:
\begin{equation}
\label{eq:wnc}
D_i=\varepsilon\frac{\delta_{i,0}}{(\Delta x)^d}\;\;\;\;\;\;
\Longrightarrow\;\;\;\;\;\;D_\mu=\varepsilon
\end{equation}
and the Fokker-Planck equation is
\begin{eqnarray}
\label{eq:fpemul3}
\frac{\partial P}{\partial t}= -\sum_\mu \frac{\partial}
{\partial u_\mu} f_\mu P
\;\;\;\;\;\;\;\;\;\;\;\;\;\;\;\;\;\;\;\;\;\;\;\;\;\;
\;\;\;\;\;\;\;\;\;\;\;\;\;\;
\nonumber \\
+ \frac{\varepsilon}{(L\Delta x)^{d}}
\;\sum_{\mu,\nu,\eta} \;\frac{\partial}{\partial u_\mu}
\;g_{\mu-\nu}\; \frac{\partial}{\partial u_\eta}
\;g_{\eta+\nu}\;P
\end{eqnarray}

\vskip1cm
\noindent
{\em $^*$ Permanent address:
Dept. de F\'{\i}sica i Enginyeria
Nuclear, E.T.S. d'Enginyers Industrials
de Terrassa, Univ. Polit\`{e}cnica de Catalunya,
Colom 11, E-08222 Terrassa, Spain.}

\end{document}